\documentclass[twocolumn,showpacs,aps,prl,nofootinbib]{revtex4}
\usepackage{multirow}
\usepackage{amsmath}
\usepackage{epsfig}
\usepackage{subfigure}
\usepackage{float}
\usepackage{verbatim} 
\usepackage{booktabs}

\begin{document}

\bigskip
\title{Next-to-leading-order predictions for $t$-channel single-top production at hadron colliders}

\author{J.~M.~Campbell$^{a}$, R.~Frederix$^{b,c}$, F.~Maltoni$^{c}$, F.~Tramontano$^{d}$}
\affiliation{
$^{a}$Department of Physics and Astronomy, University of Glasgow, 
Glasgow G12 8QQ, United Kingdom \\
$^{b}$ PH Department, Theory group, CERN 
1211-CH Geneva, Switzerland\\
$^{c}$Center for Particle Physics and Phenomenology (CP3), Universit\'e catholique de Louvain, B-1348 
Louvain-la-Neuve, Belgium\\
$^{d}$ Universit\`a di Napoli Federico II, Dipartimento di Scienze Fisiche, and INFN, Sezione di Napoli, I-80126 Napoli, Italy
}

\begin{abstract}
  We present the predictions at next-to-leading order (NLO) in the
  strong coupling for the single-top cross section in the $t$ channel
  at the Tevatron and the LHC. Our calculation starts from the $2\to
  3$ Born amplitude $g q \to t \bar b q'$, keeping the $b$-quark mass
  non-zero.  A comparison is performed with a traditional NLO
  calculation of this channel based on the $2\to2$ Born process with a
  bottom quark in the initial state.  In particular, the effect of
  using kinematic approximations and resumming logarithms of the form
  $\log(Q^2/m_b^2)$ in the $2\to2$ process is assessed.  Our results
  show that the $2\to 3$ calculation is very well behaved and in
  substantial agreement with the predictions based on the $2 \to 2$
  process.
\end{abstract}


\maketitle


It is a quite remarkable fact that in hadron collisions, top quarks
can be produced via electroweak interactions at a rate comparable with
strong production~\cite{Willenbrock:1986cr,Yuan:1989tc,Ellis:1992yw}.
Such unique behavior is mainly due to two factors. First, a top quark
can be produced together with its $SU(2)_L$ partner, the bottom quark,
with a sizable gain in phase space cost with respect to a top and
anti-top quark pair. Second, among the three possible production
channels, one entails the exchange of a vector boson in the
$t$-channel, leading to an enhanced cross section at high energies.

Given the large predicted cross section, evidence for single top
production has been actively sought and recently established at the
Tevatron~\cite{Abazov:2006gd,Aaltonen:2008sy} and it will play an
important role in the physics program at the LHC.  Single-top
production offers, for instance, the only effective way of extracting
direct information on $V_{tb}$~\cite{Alwall:2006bx}.  In fact, at the
Tevatron the prospects for the detection and then measurement of the
electroweak (EW) production cross sections have significantly worsened since the
first theoretical proposals~\cite{Stelzer:1998ni}.  The main reason
for this was an underestimate of the impact of large backgrounds such
as those coming from $W+$ jet production (both with and without heavy
flavors) and from the strong production of $t \bar
t$~\cite{Bowen:2004my}.  The situation at the LHC, though bound to
improve thanks to the larger rates expected, will not be qualitatively
very different.

The most accurate analyses for single top are based on two essential
ingredients.  The first is an in situ determination of the background
rates. Predictions from theory are in this case not able to match the
needed accuracy. The second is the systematic exploitation of
theoretical predictions for the kinematic properties of signal (and
backgrounds). This information is encoded via sophisticated analysis
techniques (such as those based on matrix elements, neural networks
and others~\cite{Abazov:2006gd,Aaltonen:2008sy}). Such methods are
crucial in building efficient discriminating variables to select the
Standard Model signal or possibly find indications of new physics
effects~\cite{Tait:2000sh}.

It is therefore clear that the most accurate predictions for the
signal, both for rates and kinematic distributions, are needed as
inputs in these analyses.  An intense activity in the last fifteen
years has led to increasingly-sophisticated predictions at NLO
accuracy.  Calculations have progressed from evaluations of total
rates~\cite{Bordes:1994ki,Stelzer:1997ns}, to differential
distributions~\cite{Harris:2002md,Kidonakis:2006bu}, including 
spin correlations in production and 
decay~\cite{Campbell:2004ch,Campbell:2005bb,Cao:2004ap,Cao:2005pq} 
and finally to the implementation of the three production channels in a
fully exclusive Monte Carlo program~\cite{Frixione:2005vw,Frixione:2008yi}.

\begin{figure}[ht!]
\centering
\subfigure[]{\includegraphics[scale=.6]{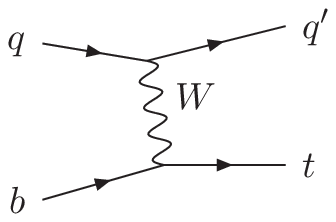}}
\hspace*{.2cm}
\subfigure[]{\includegraphics[scale=.6]{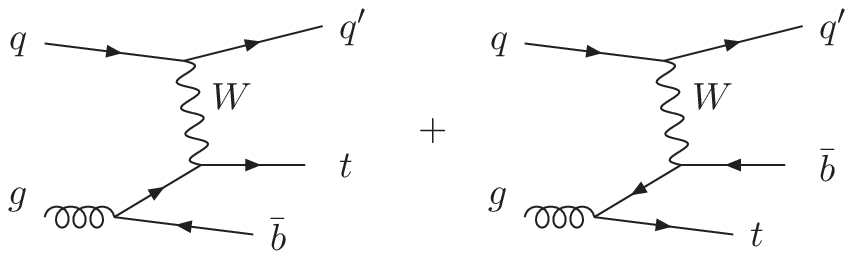}}
\caption{Diagrams contributing at LO in the $2\to 2$ (a) and $2 \to
  3$ (b) approaches.}
\label{diagrams}
\end{figure}

All NLO calculations available so far are based on the $2\to 2$
scattering process, Fig.~\ref{diagrams}(a), where a $b$-quark appears
in the initial state~\cite{Collins:1998rz,Kramer:2000hn}. The
usefulness of such an approach, called the five-flavor (5F) scheme, is
twofold. Firstly, the calculation greatly simplifies (as we shall
describe in detail later), leading to straightforward calculations and
compact results.  Second, possibly large logarithms of the form
$\log{(Q^2/m_b^2)}$ due to initial state collinear configurations with
$g\to b\bar b$ splitting are consistently resummed into the $b$-quark
parton distribution functions leading to an improved stability of the
perturbative expansion. Effects related to the ``spectator $b$'', such
as the presence of a $b$-jet and the $b$ mass, Fig.~\ref{diagrams}(b),
only enter at NLO. As a result, most of the current calculations and
corresponding Monte Carlo
implementations~\cite{Harris:2002md,Campbell:2004ch,Campbell:2005bb,Cao:2004ap,Cao:2005pq,Frixione:2005vw,Frixione:2008yi}
do not accurately model such effects.  An alternative approach is to
consider as Born the $2\to 3$ scattering process,
Fig.~\ref{diagrams}(b), keeping a finite $b$ mass. In this scheme,
called the four-flavor (4F) scheme, the $b$ quarks do not enter in the
QCD evolution of the PDF's and of the strong coupling. The calculation
of the NLO corrections is much more involved due to the inclusion of
an additional parton in the final state and the presence of a further
mass scale. However, features associated with the kinematic
description of the spectator $b$'s can be genuinely investigated at
NLO accuracy.  In this Letter we present NLO results in the 4F scheme
and compare them to those in the 5F scheme. The two approaches, being
by definition equivalent, would give the same results at all orders in
the perturbative expansion. At fixed (low) order, however, predictions
could in principle differ significantly and the question of the range
of applicability of each approach is raised.

\begin{figure}[ht!]
\centering
\subfigure[]{\includegraphics[scale=.6]{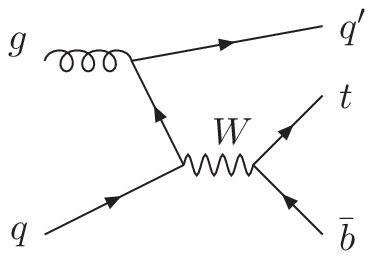}}
\hspace*{.2cm}
\subfigure[]{\includegraphics[scale=.6]{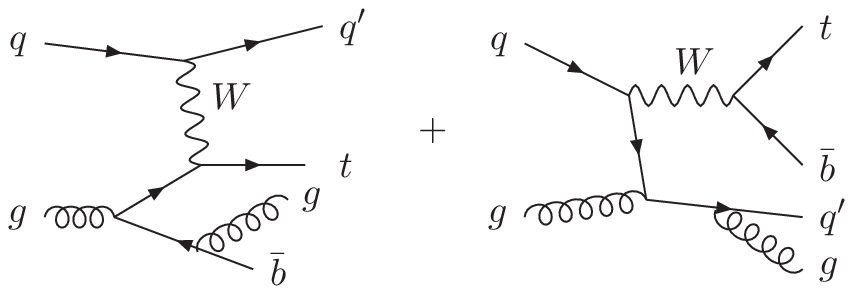}}
\caption{(a) Representative $s$-channel diagram present in $gq\to t
  \bar b q'$.  Interference of these diagrams with those of
  Fig.~\ref{diagrams}(b) vanishes because of color.  (b)
  Representative $t$- and $s$-channel diagrams for $gq\to t \bar b
  gq'$: their interference is suppressed by $1/N_c^2$. }
\label{diagrams2}
\end{figure}

A very important simplification in the $2 \to 2$ calculation at NLO is
that QCD corrections completely factorize in terms of a light and
heavy currents: color conservation forbids the interference between
diagrams (1-loop/Born or real/real) where the light quark and the
heavy quark lines are connected by a gluon. In addition, real/real
interferences between $t$-channel (Fig.~\ref{diagrams}(b)) and
$s$-channel (Fig.~\ref{diagrams2}(a)) diagrams vanish.  A clear
separation between the two processes is therefore maintained at NLO.

Quite remarkably, QCD corrections to the $2\to 3$ Born still display
almost a complete factorization.  Most of the real emission processes
can be uniquely associated ({\it i.e.}, in a gauge invariant way) with
either the light quark current or the heavy quark
current. Interference terms are either exactly zero or
color-suppressed by $1/N_c^2$. We have kept the color suppressed
interferences between light and heavy currents, but not those between
$s$-channel NNLO real corrections and the $t$-channel real corrections
(depicted in Fig.~\ref{diagrams2}(b)). Interferences in
  $q\bar q \to t \bar b q' \bar q$ subprocess between
  $t$-channel diagrams and those with an on-shell $W$, $q\bar q \to t
  \bar b (W^-\to q \bar q')$ vanish in the $W$ zero-width limit and have
  not been included.  However, we have checked that all the neglected
interferences are very small and de facto do not hamper a meaningful
separation of the channels.

All the analytic calculations presented in this paper have been
performed with FORM~\cite{Vermaseren:2000nd}: tree-level and loop
matrix elements are computed at the helicity amplitude level and
therefore top spin information is available. Tree-level matrix
elements and the LO results for cross sections and distributions have
been accurately checked with
MadGraph/MadEvent~\cite{Alwall:2007st}. The loop contributions have
been evaluated in both the dimensional reduction and four dimensional
helicity schemes, following the procedure outlined in
Ref.~\cite{Harris:2002md}. Tensor integrals have been decomposed with
the help of a reduction routine based on the Passarino-Veltman
approach~\cite{Passarino:1978jh}. Scalar integrals have been
explicitly computed with standard methods and compared numerically
with those of Ref.~\cite{Ellis:2007qk}.  Gauge invariance, CP, and
kinematic symmetries ($m_b \leftrightarrow m_t$) have been extensively
used to check the consistency of the calculation. Infrared and
collinear divergences in the integrated real and virtual contributions
have been cancelled locally through the use of the dipole subtraction
technique~\cite{Catani:1996vz} in its massive
formulation~\cite{Catani:2002hc} as implemented in
MCFM~\cite{Campbell:2000bg}.  The contributions from the dipole
counterterms have been generated independently by
MadDipole~\cite{Frederix:2008hu} and checked point-by-point in phase
space.  Finally, as the most important check of our calculation, we
have derived the results for $e^+ e^- \to Z/\gamma^\star \to b \bar b
g$ at NLO. This calculation can be obtained from ours by simply
accounting for the difference in the weak couplings, by setting
$m_b=m_t$, by ignoring the QCD corrections on the light current and by
setting the $W$ virtuality positive. We found excellent agreement with
the results of Ref.~\cite{Nason:1997nw}, that have been obtained in
completely different subtraction and regularization schemes.


\begin{figure}[t]
\centering
\includegraphics[scale=.6]{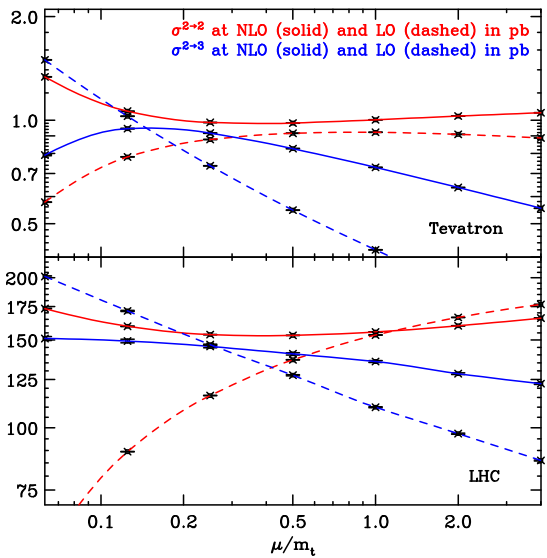}
\caption{Scale dependence of the $2\to 2$ and $2 \to 3$ calculations,
  at LO (dashed) and NLO (solid) order.  Factorization and
  renormalization scales in the heavy and light quark lines are equal
  to $\mu$.  For the LHC only top production is considered, the
  behaviour of the anti-top being very similar.}
\label{fig:scale}
\end{figure}

We now present and discuss the results of the NLO calculation in the 4F
scheme (Born $2\to3$) and compare with those of the 5F scheme (Born
$2\to2$), at the Tevatron ($p\bar p$, $\sqrt{s}=1.96$ TeV) and the LHC
($p p$, $\sqrt{s}=14$ TeV). In our studies we assume $m_t=172$~GeV,
$m_b=4.7$~GeV and use the CTEQ6.6 PDF set~\cite{Nadolsky:2008zw}.  
For the $2\to3$ calculation we  pass to the 4F scheme by adding suitable
finite terms, as explained in Ref.~\cite{Cacciari:1998it}. As an independent check 
we have verified that results obtained with the explicit four-flavor 
MRST set~\cite{Martin:2006qz} are fully consistent  with  those obtained 
in the corresponding five-flavor MRST set plus the finite terms. 
In order to perform a fair comparison between the two schemes, we strictly 
follow the approach of Refs.~\cite{Collins:1998rz,Kramer:2000hn} and, contrary to most of the
available MC implementations~\cite{Harris:2002md,Campbell:2004ch,Campbell:2005bb,Cao:2004ap,Cao:2005pq,Frixione:2005vw,Frixione:2008yi}
in the 5F scheme, we compute the real diagrams $g q \to t \bar b q'$
with a non-zero $b$ mass. We have checked that this has a negligible
effect on the total cross sections, as well as in the top and light
jet distributions at the Tevatron~\cite{Sullivan:2004ie} and the
LHC. On the other hand, the distributions of the spectator $b$'s are
significantly affected.

In Fig.~\ref{fig:scale} we show the cross sections for top production
at the Tevatron and the LHC in the two schemes as a function of
$\mu/m_t$, where $\mu$ is a common renormalization and factorization
scale.  The 4F calculation has a stronger dependence on the scale than
the 5F one, particularly at the Tevatron, which simply reflects the
fact that the $2\to3$ Born calculation already contains a factor of
$\alpha_s$.  However, we observe that both calculations are much more
stable under scale variations at NLO than at LO.  To establish an
optimal central value for the scales, we have studied separately the
scale dependence associated with the light and heavy quark lines.  As
expected, most of the overall scale dependence is inherited from the
heavy quark line.  In the 4F scheme it is minimal for scales around
$m_t/2$ and $m_t/4$ for the light and heavy quark lines respectively,
which therefore sets our central scale choice. In the 5F scheme the
scale dependence is very mild and we simply choose $m_t$ for both
lines.

\begin{table}[b]
\begin{small}
\renewcommand{\arraystretch}{1.3}
\begin{center}
\begin{tabular}{c@{}r@{ }lr@{ }lr@{ }l}
\toprule[0.08em]
\multirow{2}*{Born}&\multicolumn{2}{c}{TeV $t\,(=\bar{t})$}&\multicolumn{2}{c}{LHC $t$}&\multicolumn{2}{c}{LHC $\bar{t}$}\\[-2pt]
&$\qquad$(LO)&NLO&(LO)&NLO&(LO)&NLO\\
\midrule[0.05em]
$2 \to 2$ &(0.92)&$1.00^{+0.03+0.10}_{-0.02-0.08}$&(153)&$156^{+4+3}_{-4-4}$&(89)&$93^{+3+2}_{-2-2}$\\
$2 \to 3$ &(0.68)&$0.94^{+0.07+0.08}_{-0.11-0.07}$&(143)&$146^{+4+3}_{-7-3}$&(81)&$86^{+4+2}_{-3-2}$\\
\bottomrule[0.08em]
\end{tabular}
\end{center}
\end{small}
\caption{Inclusive cross sections (in pb) for $t$-channel single top
  production at the Tevatron and LHC using (CTEQ6L1)
  CTEQ6.6 PDF's for the (LO) NLO predictions and
  $\mu_0^L=m_t$ ($\mu_0^H=m_t$) and $\mu_0^L=m_t/2$ ($\mu_0^H=m_t/4$)
  as central values for the factorization and renormalization scales
  for the light (heavy) line in the 5F and 4F schemes, respectively.
  The first uncertainty comes from scale variations, the second
  from PDF errors.
  \label{tab:one}
}
\end{table}

Table~\ref{tab:one} shows the predictions for the total cross sections
in the two schemes, together with their uncertainties. The scale
uncertainties are evaluated by varying the renormalization and
factorization scales independently between $\mu_0^{L,H}/2 < \mu_{F,R}
< 2\mu_0^{L,H}$ with $1/2 < \mu_F/\mu_R <2$ and $\mu^L/\mu^H$
constant.  We see that the uncertainty in the 4F scheme is larger than
(similar to) that in the 5F scheme at the Tevatron (LHC).  The
difference between the NLO predictions in the two schemes is rather
small, with uncertainties typically less than 5\% in both cases. The
exception is the 4F calculation at the Tevatron with an uncertainty of
around 10\%, which is however still of the same order as the absolute
difference with the 5F calculation.  The small scale uncertainties
together with quite modest increases of the cross sections from LO to
NLO provide a clear indication that the perturbative expansions are
very well behaved.

\begin{figure}[t]
\centering
\includegraphics[scale=.6]{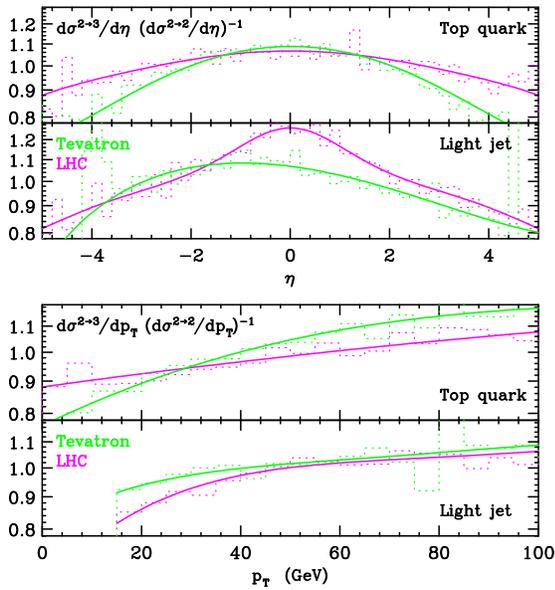}
\caption{Shape comparison for the top quark and light jet NLO
distributions. The bin-by-bin ratio of the normalized (4F and 5F) 
distributions  in $\eta$ and $p_T$  is shown.
}
\label{fig:ratios}
\end{figure}

\begin{figure}[t]
\centering
\includegraphics[scale=.6]{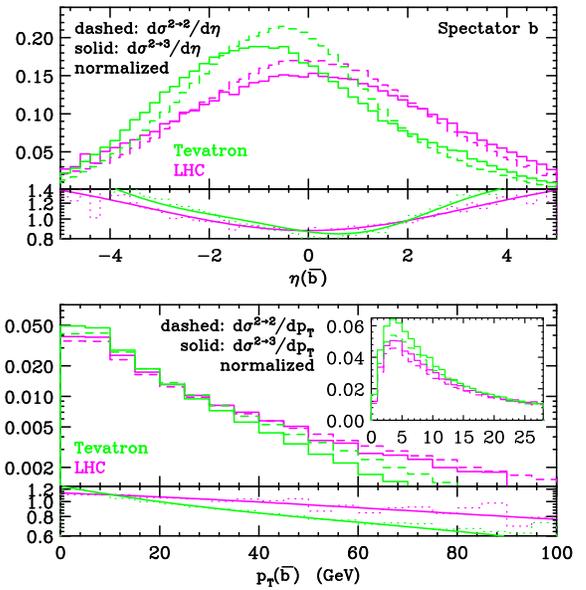}
\caption{Distributions (upper plots) in $\eta$ and $p_T$ and shape
  comparison (lower plots) of the $2 \to 3$ and $2 \to 2$ calculations
  of the spectator $b$ at NLO.}
\label{fig:b}
\end{figure}

In Fig.~\ref{fig:ratios} we compare NLO predictions for the top quark
and light jet pseudo rapidity $\eta$ and transverse momentum $p_T$.
To define the light jet we used the $k_T$ algorithm and imposed
$p_T>15$ GeV, $\Delta R>0.7$. Results are presented as a bin-by-bin
ratio of the normalized (4F and 5F) distributions.  For the LHC only
top production is shown, with the behaviour of the anti-top very
similar.  Although the predictions differ somewhat, the differences
are typically at the 10\% level and always less than 20\%.  Finally, 
we study the NLO distributions in $\eta$ and $p_T$ for the spectator $b$. 
We find that the fraction of events at the Tevatron (LHC) 
where the $b$ is central and at high-$p_T$ ($|\eta|<2.5, p_T>20$ GeV)  is  28\% (36\%) 
with a very small scale dependence. From Fig.~\ref{fig:b} we see that the largest effects in the shapes are present
at the Tevatron, where the spectator $b$ tends to be more forward and softer at high $p_T$ 
than in the 5F calculation (where these observables are effectively only at LO). 

We have reported on the computation of the NLO corrections to the EW
production of top and bottom quarks through the $t$-channel exchange
of a $W$ boson, keeping the mass of the heavy quarks finite. This
allows a systematic study of the approximations and improvements
associated with the different schemes for treating heavy flavors in
QCD. We find that the 4F calculation is well behaved: it displays a
10\% (4\%) scale uncertainty and a modest (very small) increase of the cross section 
from LO to NLO at the Tevatron (LHC).
It gives rates that are slightly smaller than the 5F predictions (by about 6\%).  The
two calculations are consistent at the Tevatron, where the uncertainty
of the 4F calculation is similar to their difference and marginally
consistent at the LHC, where the estimated uncertainties are much smaller. 
Such a difference could be interpreted as resulting from the
resummation of the $\log{(Q^2/m_b^2)}$ in the 5F calculation, or as an
indication of the need for even higher order corrections. A NNLO
prediction in the 5F scheme could help in settling this issue.  The 4F
calculation provides reliable predictions for all the relevant
differential distributions, in particular offering for the first time
genuine NLO predictions for the spectator $b$ rapidity and $p_T$.  A
detailed comparison with current Monte Carlo based predictions and an
extension of this study to the production of fourth generation quarks,
such as $t^\prime b$, $t b^\prime$ and $t^\prime b^\prime$ are left to
forthcoming studies.


We are indebted to Scott Willenbrock for numerous comments and
insights. We also thank Carlo Oleari and Paolo Nason for their
valuable help in comparing with the calculation of
Ref.~\cite{Nason:1997nw}.  This work is partially supported by the
Belgian Federal Office for Scientific, Technical and Cultural Affairs
through the Interuniversity Attraction Pole No. P6/11.

\bibliography{physics}

\begin{thebibliography}{34}
\expandafter\ifx\csname natexlab\endcsname\relax\def\natexlab#1{#1}\fi
\expandafter\ifx\csname bibnamefont\endcsname\relax
  \def\bibnamefont#1{#1}\fi
\expandafter\ifx\csname bibfnamefont\endcsname\relax
  \def\bibfnamefont#1{#1}\fi
\expandafter\ifx\csname citenamefont\endcsname\relax
  \def\citenamefont#1{#1}\fi
\expandafter\ifx\csname url\endcsname\relax
  \def\url#1{\texttt{#1}}\fi
\expandafter\ifx\csname urlprefix\endcsname\relax\def\urlprefix{URL }\fi
\providecommand{\bibinfo}[2]{#2}
\providecommand{\eprint}[2][]{\url{#2}}

\bibitem[{\citenamefont{Willenbrock and Dicus}(1986)}]{Willenbrock:1986cr}
\bibinfo{author}{\bibfnamefont{S.~S.~D.} \bibnamefont{Willenbrock}}
  \bibnamefont{and} \bibinfo{author}{\bibfnamefont{D.~A.} \bibnamefont{Dicus}},
  \bibinfo{journal}{Phys. Rev.} \textbf{\bibinfo{volume}{D34}},
  \bibinfo{pages}{155} (\bibinfo{year}{1986}).

\bibitem[{\citenamefont{Yuan}(1990)}]{Yuan:1989tc}
\bibinfo{author}{\bibfnamefont{C.~P.} \bibnamefont{Yuan}},
  \bibinfo{journal}{Phys. Rev.} \textbf{\bibinfo{volume}{D41}},
  \bibinfo{pages}{42} (\bibinfo{year}{1990}).

\bibitem[{\citenamefont{Ellis and Parke}(1992)}]{Ellis:1992yw}
\bibinfo{author}{\bibfnamefont{R.~K.} \bibnamefont{Ellis}} \bibnamefont{and}
  \bibinfo{author}{\bibfnamefont{S.~J.} \bibnamefont{Parke}},
  \bibinfo{journal}{Phys. Rev.} \textbf{\bibinfo{volume}{D46}},
  \bibinfo{pages}{3785} (\bibinfo{year}{1992}).

\bibitem[{\citenamefont{Abazov et~al.}(2007)}]{Abazov:2006gd}
\bibinfo{author}{\bibfnamefont{V.~M.} \bibnamefont{Abazov}}
  \bibnamefont{et~al.} (\bibinfo{collaboration}{D0}), \bibinfo{journal}{Phys.
  Rev. Lett.} \textbf{\bibinfo{volume}{98}}, \bibinfo{pages}{181802}
  (\bibinfo{year}{2007}).

\bibitem[{\citenamefont{Aaltonen et~al.}(2008)}]{Aaltonen:2008sy}
\bibinfo{author}{\bibfnamefont{T.}~\bibnamefont{Aaltonen}} \bibnamefont{et~al.}
  (\bibinfo{collaboration}{CDF}), \bibinfo{journal}{Phys. Rev. Lett.}
  \textbf{\bibinfo{volume}{101}}, \bibinfo{pages}{252001}
  (\bibinfo{year}{2008}).

\bibitem[{\citenamefont{Alwall et~al.}(2007{\natexlab{a}})}]{Alwall:2006bx}
\bibinfo{author}{\bibfnamefont{J.}~\bibnamefont{Alwall}} \bibnamefont{et~al.},
  \bibinfo{journal}{Eur. Phys. J.} \textbf{\bibinfo{volume}{C49}},
  \bibinfo{pages}{791} (\bibinfo{year}{2007}{\natexlab{a}}).

\bibitem[{\citenamefont{Stelzer et~al.}(1998)\citenamefont{Stelzer, Sullivan,
  and Willenbrock}}]{Stelzer:1998ni}
\bibinfo{author}{\bibfnamefont{T.}~\bibnamefont{Stelzer}},
  \bibinfo{author}{\bibfnamefont{Z.}~\bibnamefont{Sullivan}}, \bibnamefont{and}
  \bibinfo{author}{\bibfnamefont{S.}~\bibnamefont{Willenbrock}},
  \bibinfo{journal}{Phys. Rev.} \textbf{\bibinfo{volume}{D58}},
  \bibinfo{pages}{094021} (\bibinfo{year}{1998}).

\bibitem[{\citenamefont{Bowen et~al.}(2005)\citenamefont{Bowen, Ellis, and
  Strassler}}]{Bowen:2004my}
\bibinfo{author}{\bibfnamefont{M.~T.} \bibnamefont{Bowen}},
  \bibinfo{author}{\bibfnamefont{S.~D.} \bibnamefont{Ellis}}, \bibnamefont{and}
  \bibinfo{author}{\bibfnamefont{M.~J.} \bibnamefont{Strassler}},
  \bibinfo{journal}{Phys. Rev.} \textbf{\bibinfo{volume}{D72}},
  \bibinfo{pages}{074016} (\bibinfo{year}{2005}).

\bibitem[{\citenamefont{Tait and Yuan}(2001)}]{Tait:2000sh}
\bibinfo{author}{\bibfnamefont{T.~M.~P.} \bibnamefont{Tait}} \bibnamefont{and}
  \bibinfo{author}{\bibfnamefont{C.~P.} \bibnamefont{Yuan}},
  \bibinfo{journal}{Phys. Rev.} \textbf{\bibinfo{volume}{D63}},
  \bibinfo{pages}{014018} (\bibinfo{year}{2001}).

\bibitem[{\citenamefont{Bordes and van Eijk}(1995)}]{Bordes:1994ki}
\bibinfo{author}{\bibfnamefont{G.}~\bibnamefont{Bordes}} \bibnamefont{and}
  \bibinfo{author}{\bibfnamefont{B.}~\bibnamefont{van Eijk}},
  \bibinfo{journal}{Nucl. Phys.} \textbf{\bibinfo{volume}{B435}},
  \bibinfo{pages}{23} (\bibinfo{year}{1995}).

\bibitem[{\citenamefont{Stelzer et~al.}(1997)\citenamefont{Stelzer, Sullivan,
  and Willenbrock}}]{Stelzer:1997ns}
\bibinfo{author}{\bibfnamefont{T.}~\bibnamefont{Stelzer}},
  \bibinfo{author}{\bibfnamefont{Z.}~\bibnamefont{Sullivan}}, \bibnamefont{and}
  \bibinfo{author}{\bibfnamefont{S.}~\bibnamefont{Willenbrock}},
  \bibinfo{journal}{Phys. Rev.} \textbf{\bibinfo{volume}{D56}},
  \bibinfo{pages}{5919} (\bibinfo{year}{1997}).

\bibitem[{\citenamefont{Harris et~al.}(2002)\citenamefont{Harris, Laenen, Phaf,
  Sullivan, and Weinzierl}}]{Harris:2002md}
\bibinfo{author}{\bibfnamefont{B.~W.} \bibnamefont{Harris}},
  \bibinfo{author}{\bibfnamefont{E.}~\bibnamefont{Laenen}},
  \bibinfo{author}{\bibfnamefont{L.}~\bibnamefont{Phaf}},
  \bibinfo{author}{\bibfnamefont{Z.}~\bibnamefont{Sullivan}}, \bibnamefont{and}
  \bibinfo{author}{\bibfnamefont{S.}~\bibnamefont{Weinzierl}},
  \bibinfo{journal}{Phys. Rev.} \textbf{\bibinfo{volume}{D66}},
  \bibinfo{pages}{054024} (\bibinfo{year}{2002}).

\bibitem[{\citenamefont{Kidonakis}(2006)}]{Kidonakis:2006bu}
\bibinfo{author}{\bibfnamefont{N.}~\bibnamefont{Kidonakis}},
  \bibinfo{journal}{Phys. Rev.} \textbf{\bibinfo{volume}{D74}},
  \bibinfo{pages}{114012} (\bibinfo{year}{2006}).

\bibitem[{\citenamefont{Campbell et~al.}(2004)\citenamefont{Campbell, Ellis,
  and Tramontano}}]{Campbell:2004ch}
\bibinfo{author}{\bibfnamefont{J.~M.} \bibnamefont{Campbell}},
  \bibinfo{author}{\bibfnamefont{R.~K.} \bibnamefont{Ellis}}, \bibnamefont{and}
  \bibinfo{author}{\bibfnamefont{F.}~\bibnamefont{Tramontano}},
  \bibinfo{journal}{Phys. Rev.} \textbf{\bibinfo{volume}{D70}},
  \bibinfo{pages}{094012} (\bibinfo{year}{2004}).

\bibitem[{\citenamefont{Campbell and Tramontano}(2005)}]{Campbell:2005bb}
\bibinfo{author}{\bibfnamefont{J.~M.} \bibnamefont{Campbell}} \bibnamefont{and}
  \bibinfo{author}{\bibfnamefont{F.}~\bibnamefont{Tramontano}},
  \bibinfo{journal}{Nucl. Phys.} \textbf{\bibinfo{volume}{B726}},
  \bibinfo{pages}{109} (\bibinfo{year}{2005}).

\bibitem[{\citenamefont{Cao et~al.}(2005{\natexlab{a}})\citenamefont{Cao,
  Schwienhorst, and Yuan}}]{Cao:2004ap}
\bibinfo{author}{\bibfnamefont{Q.-H.} \bibnamefont{Cao}},
  \bibinfo{author}{\bibfnamefont{R.}~\bibnamefont{Schwienhorst}},
  \bibnamefont{and} \bibinfo{author}{\bibfnamefont{C.~P.} \bibnamefont{Yuan}},
  \bibinfo{journal}{Phys. Rev.} \textbf{\bibinfo{volume}{D71}},
  \bibinfo{pages}{054023} (\bibinfo{year}{2005}{\natexlab{a}}).

\bibitem[{\citenamefont{Cao et~al.}(2005{\natexlab{b}})\citenamefont{Cao,
  Schwienhorst, Benitez, Brock, and Yuan}}]{Cao:2005pq}
\bibinfo{author}{\bibfnamefont{Q.-H.} \bibnamefont{Cao}},
  \bibinfo{author}{\bibfnamefont{R.}~\bibnamefont{Schwienhorst}},
  \bibinfo{author}{\bibfnamefont{J.~A.} \bibnamefont{Benitez}},
  \bibinfo{author}{\bibfnamefont{R.}~\bibnamefont{Brock}}, \bibnamefont{and}
  \bibinfo{author}{\bibfnamefont{C.~P.} \bibnamefont{Yuan}},
  \bibinfo{journal}{Phys. Rev.} \textbf{\bibinfo{volume}{D72}},
  \bibinfo{pages}{094027} (\bibinfo{year}{2005}{\natexlab{b}}).

\bibitem[{\citenamefont{Frixione et~al.}(2006)\citenamefont{Frixione, Laenen,
  Motylinski, and Webber}}]{Frixione:2005vw}
\bibinfo{author}{\bibfnamefont{S.}~\bibnamefont{Frixione}},
  \bibinfo{author}{\bibfnamefont{E.}~\bibnamefont{Laenen}},
  \bibinfo{author}{\bibfnamefont{P.}~\bibnamefont{Motylinski}},
  \bibnamefont{and} \bibinfo{author}{\bibfnamefont{B.~R.}
  \bibnamefont{Webber}}, \bibinfo{journal}{JHEP} \textbf{\bibinfo{volume}{03}},
  \bibinfo{pages}{092} (\bibinfo{year}{2006}).

\bibitem[{\citenamefont{Frixione et~al.}(2008)\citenamefont{Frixione, Laenen,
  Motylinski, Webber, and White}}]{Frixione:2008yi}
\bibinfo{author}{\bibfnamefont{S.}~\bibnamefont{Frixione}},
  \bibinfo{author}{\bibfnamefont{E.}~\bibnamefont{Laenen}},
  \bibinfo{author}{\bibfnamefont{P.}~\bibnamefont{Motylinski}},
  \bibinfo{author}{\bibfnamefont{B.~R.} \bibnamefont{Webber}},
  \bibnamefont{and} \bibinfo{author}{\bibfnamefont{C.~D.} \bibnamefont{White}},
  \bibinfo{journal}{JHEP} \textbf{\bibinfo{volume}{07}}, \bibinfo{pages}{029}
  (\bibinfo{year}{2008}).

\bibitem[{\citenamefont{Collins}(1998)}]{Collins:1998rz}
\bibinfo{author}{\bibfnamefont{J.~C.} \bibnamefont{Collins}},
  \bibinfo{journal}{Phys. Rev.} \textbf{\bibinfo{volume}{D58}},
  \bibinfo{pages}{094002} (\bibinfo{year}{1998}).

\bibitem[{\citenamefont{Kramer et~al.}(2000)\citenamefont{Kramer, Olness, and
  Soper}}]{Kramer:2000hn}
\bibinfo{author}{\bibfnamefont{M.}~\bibnamefont{Kramer}},
  \bibinfo{author}{\bibfnamefont{F.~I.} \bibnamefont{Olness}},
  \bibnamefont{and} \bibinfo{author}{\bibfnamefont{D.~E.} \bibnamefont{Soper}},
  \bibinfo{journal}{Phys. Rev.} \textbf{\bibinfo{volume}{D62}},
  \bibinfo{pages}{096007} (\bibinfo{year}{2000}).

\bibitem[{\citenamefont{Vermaseren}(2000)}]{Vermaseren:2000nd}
\bibinfo{author}{\bibfnamefont{J.~A.~M.} \bibnamefont{Vermaseren}}
  (\bibinfo{year}{2000}), \eprint{math-ph/0010025}.

\bibitem[{\citenamefont{Alwall et~al.}(2007{\natexlab{b}})}]{Alwall:2007st}
\bibinfo{author}{\bibfnamefont{J.}~\bibnamefont{Alwall}} \bibnamefont{et~al.},
  \bibinfo{journal}{JHEP} \textbf{\bibinfo{volume}{09}}, \bibinfo{pages}{028}
  (\bibinfo{year}{2007}{\natexlab{b}}).

\bibitem[{\citenamefont{Passarino and Veltman}(1979)}]{Passarino:1978jh}
\bibinfo{author}{\bibfnamefont{G.}~\bibnamefont{Passarino}} \bibnamefont{and}
  \bibinfo{author}{\bibfnamefont{M.~J.~G.} \bibnamefont{Veltman}},
  \bibinfo{journal}{Nucl. Phys.} \textbf{\bibinfo{volume}{B160}},
  \bibinfo{pages}{151} (\bibinfo{year}{1979}).

\bibitem[{\citenamefont{Ellis and Zanderighi}(2008)}]{Ellis:2007qk}
\bibinfo{author}{\bibfnamefont{R.~K.} \bibnamefont{Ellis}} \bibnamefont{and}
  \bibinfo{author}{\bibfnamefont{G.}~\bibnamefont{Zanderighi}},
  \bibinfo{journal}{JHEP} \textbf{\bibinfo{volume}{02}}, \bibinfo{pages}{002}
  (\bibinfo{year}{2008}).

\bibitem[{\citenamefont{Catani and Seymour}(1997)}]{Catani:1996vz}
\bibinfo{author}{\bibfnamefont{S.}~\bibnamefont{Catani}} \bibnamefont{and}
  \bibinfo{author}{\bibfnamefont{M.~H.} \bibnamefont{Seymour}},
  \bibinfo{journal}{Nucl. Phys.} \textbf{\bibinfo{volume}{B485}},
  \bibinfo{pages}{291} (\bibinfo{year}{1997}).

\bibitem[{\citenamefont{Catani et~al.}(2002)\citenamefont{Catani, Dittmaier,
  Seymour, and Trocsanyi}}]{Catani:2002hc}
\bibinfo{author}{\bibfnamefont{S.}~\bibnamefont{Catani}},
  \bibinfo{author}{\bibfnamefont{S.}~\bibnamefont{Dittmaier}},
  \bibinfo{author}{\bibfnamefont{M.~H.} \bibnamefont{Seymour}},
  \bibnamefont{and}
  \bibinfo{author}{\bibfnamefont{Z.}~\bibnamefont{Trocsanyi}},
  \bibinfo{journal}{Nucl. Phys.} \textbf{\bibinfo{volume}{B627}},
  \bibinfo{pages}{189} (\bibinfo{year}{2002}).

\bibitem[{\citenamefont{Campbell and Ellis}(2000)}]{Campbell:2000bg}
\bibinfo{author}{\bibfnamefont{J.~M.} \bibnamefont{Campbell}} \bibnamefont{and}
  \bibinfo{author}{\bibfnamefont{R.~K.} \bibnamefont{Ellis}},
  \bibinfo{journal}{Phys. Rev.} \textbf{\bibinfo{volume}{D62}},
  \bibinfo{pages}{114012} (\bibinfo{year}{2000}).

\bibitem[{\citenamefont{Frederix et~al.}(2008)\citenamefont{Frederix, Gehrmann,
  and Greiner}}]{Frederix:2008hu}
\bibinfo{author}{\bibfnamefont{R.}~\bibnamefont{Frederix}},
  \bibinfo{author}{\bibfnamefont{T.}~\bibnamefont{Gehrmann}}, \bibnamefont{and}
  \bibinfo{author}{\bibfnamefont{N.}~\bibnamefont{Greiner}},
  \bibinfo{journal}{JHEP} \textbf{\bibinfo{volume}{09}}, \bibinfo{pages}{122}
  (\bibinfo{year}{2008}).

\bibitem[{\citenamefont{Nason and Oleari}(1998)}]{Nason:1997nw}
\bibinfo{author}{\bibfnamefont{P.}~\bibnamefont{Nason}} \bibnamefont{and}
  \bibinfo{author}{\bibfnamefont{C.}~\bibnamefont{Oleari}},
  \bibinfo{journal}{Nucl. Phys.} \textbf{\bibinfo{volume}{B521}},
  \bibinfo{pages}{237} (\bibinfo{year}{1998}.

\bibitem[{\citenamefont{Nadolsky et~al.}(2008)}]{Nadolsky:2008zw}
\bibinfo{author}{\bibfnamefont{P.~M.} \bibnamefont{Nadolsky}}
  \bibnamefont{et~al.}, \bibinfo{journal}{Phys. Rev.}
  \textbf{\bibinfo{volume}{D78}}, \bibinfo{pages}{013004}
  (\bibinfo{year}{2008}).

\bibitem[{\citenamefont{Cacciari et~al.}(1998)\citenamefont{Cacciari, Greco,
  and Nason}}]{Cacciari:1998it}
\bibinfo{author}{\bibfnamefont{M.}~\bibnamefont{Cacciari}},
  \bibinfo{author}{\bibfnamefont{M.}~\bibnamefont{Greco}}, \bibnamefont{and}
  \bibinfo{author}{\bibfnamefont{P.}~\bibnamefont{Nason}},
  \bibinfo{journal}{JHEP} \textbf{\bibinfo{volume}{05}}, \bibinfo{pages}{007}
  (\bibinfo{year}{1998}).

\bibitem[{\citenamefont{Martin et~al.}(2006)\citenamefont{Martin, Stirling, and
  Thorne}}]{Martin:2006qz}
\bibinfo{author}{\bibfnamefont{A.~D.} \bibnamefont{Martin}},
  \bibinfo{author}{\bibfnamefont{W.~J.} \bibnamefont{Stirling}},
  \bibnamefont{and} \bibinfo{author}{\bibfnamefont{R.~S.}
  \bibnamefont{Thorne}}, \bibinfo{journal}{Phys. Lett.}
  \textbf{\bibinfo{volume}{B636}}, \bibinfo{pages}{259} (\bibinfo{year}{2006}).

\bibitem[{\citenamefont{Sullivan}(2004)}]{Sullivan:2004ie}
\bibinfo{author}{\bibfnamefont{Z.}~\bibnamefont{Sullivan}},
  \bibinfo{journal}{Phys. Rev.} \textbf{\bibinfo{volume}{D70}},
  \bibinfo{pages}{114012} (\bibinfo{year}{2004}).

\end{thebibliography}

\end{document}